\documentclass[preprint,preprintnumbers,amsmath,amssymb]{revtex4}
\usepackage{graphicx}% Include figure files
\usepackage{dcolumn}% Align table columns on decimal point
\usepackage{bm}% bold math

\begin{document}
\title{Modeling DNA Dynamics by Path Integrals}

\author{ Marco Zoli }
\affiliation{
School of Science and Technology - CNISM \\  Universit\`{a} di Camerino, I-62032 Camerino, Italy \\ marco.zoli@unicam.it}

\date{\today}

\begin{abstract}
Complementary strands in DNA double helix show temporary fluctuational openings which are essential to biological functions such as transcription and replication of the genetic information. Such large amplitude fluctuations, known as the breathing of DNA, are generally localized and, microscopically, are due to the breaking of the hydrogen bonds linking the base pairs (\emph{bps}).
I apply imaginary time path integral techniques  to a mesoscopic Hamiltonian which accounts for the helicoidal geometry of a short circular DNA molecule. The \emph{bps} displacements with respect to the ground state are interpreted as time dependent paths whose amplitudes are consistent with the model potential for the hydrogen bonds. The portion of the paths configuration space contributing to the partition function is determined by selecting the ensemble of paths which fulfill the second law of thermodynamics. Computations of the thermodynamics in the denaturation range show the energetic advantage for the equilibrium helicoidal geometry peculiar of B-DNA. I discuss the interplay between twisting of the double helix and anharmonic stacking along the molecule backbone suggesting an interesting relation between intrinsic nonlinear character of the microscopic interactions and molecular topology.
\end{abstract}

\maketitle

\section{Hamiltonian Model for DNA}

While it has been long recognized that the form of the DNA molecule is key to understand its biological function \cite{bates}, a considerable amount of work has been recently devoted to explain the DNA dynamics in terms of the microscopic interactions at play in the double helix.
A fully atomistic description of DNA, even of a short fragment, would represent a formidable computational task due to the huge number of degrees of freedom. Moreover, the specificity of DNA lies in its capability to store the genetic information and allow a reading of the latter through large amplitude motions which temporarily bring apart portions of the complementary strands. To describe these properties we need mesoscopic models at the scale of the base pair, the fundamental entity in the nucleotide which encodes the information. The one-dimensional Dauxois-Peyrard-Bishop (DPB) model \cite{pey2} has provided a fundamental tool for the biophysicists working in the field.

The DPB Hamiltonian for a system of $N$ \emph{bps}, with reduced mass $\mu$, assumes the pair mates separation $y_n$ (for the \emph{n-th} base pair) with respect to the ground state position  as the relevant degree of freedom.  The inter-base pair interactions are modeled by a Morse potential $V_M(y_n)$ whereas the intra-base pair stacking along the molecule backbone is described by an anharmonic potential $V_S(y_n, y_{n-1})$.

\begin{figure}
\includegraphics[height=5.0cm,width=8.5cm,angle=-90]{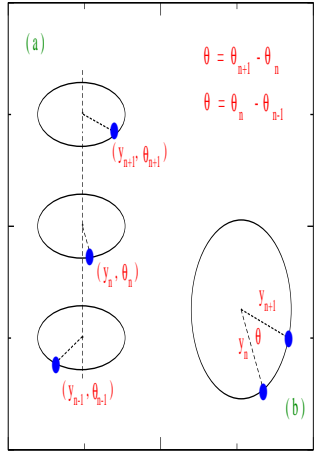}
\caption{\label{fig:1}(Color online) (a) Fixed planes picture for the right-handed helicoidal model. The blue filled circles denote the pointlike base pairs stacked along the molecule axis with twist $\theta$. The radial coordinate $y_n$ describes the $n-th$ base pair displacement from the ground state. The dashed vertical axis corresponds to the $y_n \equiv 0$ configuration, the minimum for the one-coordinate potential $V_M(y_n) + V_{sol}(y_n)$ in Eq.~(\ref{eq:1}). (b) The helix plane seen from above.}
\end{figure}

Recently, I have proposed to apply the path integral method \cite{io09} to a modified DPB Hamiltonian which includes a twist angle $\theta$ between adjacent bases, $n$ and $n-1$, along the DNA backbone \cite{io11} as shown in Fig.~\ref{fig:1}. Twisting is described by the angle that the \emph{bps} rotate around the molecule axis. B-DNA at room temperature has a helix repeat of $\sim 35 \,{\AA}$ hosting $h \sim 10$ \emph{bps}, hence the equilibrium twist angle is $\theta_{eq}= 2\pi /h \sim 0.6 \,rad$. Taking a short fragment, $N=\,100$, the (integer) equilibrium twist is $(Tw)_{eq}=\,N/h =\,10$ \cite{bates}. The Hamiltonian reads:

\begin{eqnarray}
& & H =\, \sum_{n=1}^N \biggl[ {{\mu \dot{y}_{n}^2} \over {2}} +  V_S(y_n, y_{n-1}) + V_M(y_n) + V_{sol}(y_n) \biggr] \, \nonumber
\\
& & V_S(y_n, y_{n-1})=\, {K \over 2} \Bigl[ 1 + \rho \exp\bigl[-\alpha(y_n + y_{n-1})\bigr] \Bigr] ( y_n^2 - 2 y_n y_{n-1}\cos\theta + y_{n-1}^2 ) \, \nonumber
\\
& & V_M(y_n) =\, D_n \bigl(\exp(-a_n y_n) - 1 \bigr)^2  \,
\, \nonumber
\\
& & V_{sol}(y_n) =\, - D_n f_s \bigl(\tanh(y_n/\l_s) - 1 \bigr)  \, .
\label{eq:1}
\end{eqnarray}

$D_n$ and $a_n$ are the pair dissociation energy and the inverse length setting the hydrogen bond potential range for the n-th base pair.
$K$ is the harmonic stacking whereas $\rho$ and $\alpha$ are the anharmonic stacking parameters which are taken independent of the type of base at the $n$ and $n-1$ sites.
The homogeneity assumption for the stacking relies on the observation that both types of \emph{bps} contain a purine plus a pyrimidine, the former being larger and heavier. Thus the AT- and GC- \emph{bps} are comparable in size and weight.
For the choice of the model potential parameters see Refs.\cite{io11}.
Eq.~(\ref{eq:1}) also introduces the solvent potential $V_{sol}$ which adds to $V_{M}$ thus enhancing by  $f_s D_n$ the height of the energy barrier above which the base pair dissociates.
The factor $f_s=\,0.3$ mimics the effect of a high salt concentration  which screens the negatively charged phosphate groups.
The length $l_s=\,3 {\AA}$ defines the range beyond which the Morse plateau is recovered and $D_n$ returns to be the fundamental energy scale.
For  $y_n > l_s$, the two strands are apart from each other and the hydrogen bond with the solvent is established.

\section{Path Integral Method}

The imaginary time path integral method \cite{fehi} is applied to Eq.~(\ref{eq:1}) by introducing the idea that the $y_n$ can be described by paths $x(\tau_i)$, the latter being periodic functions of the imaginary time $\tau_i$, $x(\tau_i)=\,x(\tau_i + \beta)$ with $\beta$ being the inverse temperature,. The index $i$ numbers the \emph{bps} along the $\tau$-axis. In fact, there are  $N + 1$ base pairs in Eq.~(\ref{eq:1}) but the presence of an extra base pair $y_0$ is remedied by taking periodic boundary conditions, $y_0 = \, y_N$, which close the finite chain into a loop.  This condition is  incorporated in the path integral description as the path is a closed trajectory, $x(0)=\,x(\beta)$. Hence a molecule configuration is given by $N$ paths and, in the discrete time lattice, the separation between nearest neighbors \emph{bps} is $\Delta \tau =\,\beta / N$. Then, Eq.~(\ref{eq:1}) transforms onto the time axis by mapping: $y_n \rightarrow x(\tau_i)$ and $y_{n-1} \rightarrow x(\tau_i - \Delta \tau)$.
The paths can be expanded in Fourier series with cutoff $M_F$

\begin{eqnarray}
& &x(\tau_i)=\, x_0 + \sum_{m=1}^{M_F}\Bigl[a_m \cos({2 m \pi} \tau_i / {\beta}) + b_m \sin({2 m \pi} \tau_i / {\beta}) \Bigr] \,,
\label{eq:6a}
\end{eqnarray}

and this introduces the following physical picture:

\emph{a)} given a set of coefficients $\{x_0 , a_m , b_m\}$, the $N$ \emph{bps} are represented by the configuration $\{x(\tau_i), \, i =\,1\,,..,\,N \}$.

\emph{b)} A set of coefficients corresponds to a point in the path configuration space thus, sampling the latter amounts to build an ensemble of distinct configurations for the system. As this is done for any temperature, we have a tool to describe the base pair thermal fluctuations around the equilibrium ($x(\tau_i) \sim 0$).

\emph{c)} In principle the configurations ensemble for the DNA fragment is infinite as it may include any possible combination of Fourier coefficients. For practical purposes some physical criteria intervene to select computationally the path coefficients defining a molecule configuration and contributing to the partition function. This poses a restriction on the ensemble size.

Such criteria are of two types: \emph{First}, $V_M$ excludes too negative base pair stretchings due to the hard core which mimics the repulsion between negatively charged sugar-phosphate groups.  \emph{Second}, the ensemble of paths has to be consistent with the thermodynamics laws. This means that the numerical code selects, at any temperature, a path ensemble and evaluates the entropy of the DNA fragment. If the entropy is growing versus $T$, the code proceeds to the next temperature step otherwise a new partition is performed in the Fourier coefficients integration, a new path ensemble is selected and the entropy is recalculated. This is done at any $T$ until the macroscopic constraint of the second law of thermodynamics is fulfilled throughout the whole investigated temperature range. I emphasize that the method does not put any constraint on the shape of the \emph{entropy versus $T$}- plot aside from the requirement that the entropy derivative has to be positive.
It follows that the path ensemble is a dynamical object accounting for the manyfold of molecule configurations which enter the thermodynamical calculation. The size of the ensemble is a measure of the cooperativity degree of the system. By increasing $T$, some \emph{bps} may open and cooperatively lead to bubble formation along segments of the double helix. Accordingly the ensemble size is expected to grow versus $T$.

Applying the mapping technique to Eq.~(\ref{eq:1}), the classical partition function for the DNA molecule in the solvent is

\begin{eqnarray}
& &Z_C=\oint {D}x\exp\Biggl[- \beta \sum_{i=\,1}^{N} \Bigl[{\mu \over 2}\dot{x}(\tau_i)^2 + V_S(x(\tau_i),x(\tau_i - \Delta \tau)) + V_M(x(\tau_i)) + V_{sol}(x(\tau_i)) \Bigr] \Biggr]\, \nonumber
\\
& &\oint {D}x\equiv {1 \over {\sqrt{2}\lambda_\mu}}\int dx_0 \prod_{m=1}^{M_F}\Bigl({{m \pi} \over {\lambda_\mu}}\Bigr)^2 \int da_m \int db_m \, \, ,
\label{eq:6c}
\end{eqnarray}

where  ${\lambda_\mu}=\,\sqrt{{\pi } / {\beta K}}$ is the thermal wavelength.
From Eq.~(\ref{eq:6c}), I compute the ensemble average for the displacement of the $i-th$ base pair, $< x(\tau_i) >$, which permits to get the fraction $f$ of open \emph{bps}.
As the UV signal changes quite abruptly when the \emph{bps} dissociate, $f$ is defined in terms of the Heaviside function $\vartheta(\bullet)$ as:  $f =\, {N^{-1}}\sum_{i=1}^{N} \vartheta\bigl(< x(\tau_i) > - \zeta \bigr)$.

The {\it threshold} $\zeta$ yields a criterion to establish whether an average base pair displacement is open, $<x(\tau_i)>\, \geq\, \zeta$, or not.

\begin{figure}
\includegraphics[height=5.5cm,width=8.5cm,angle=-90]{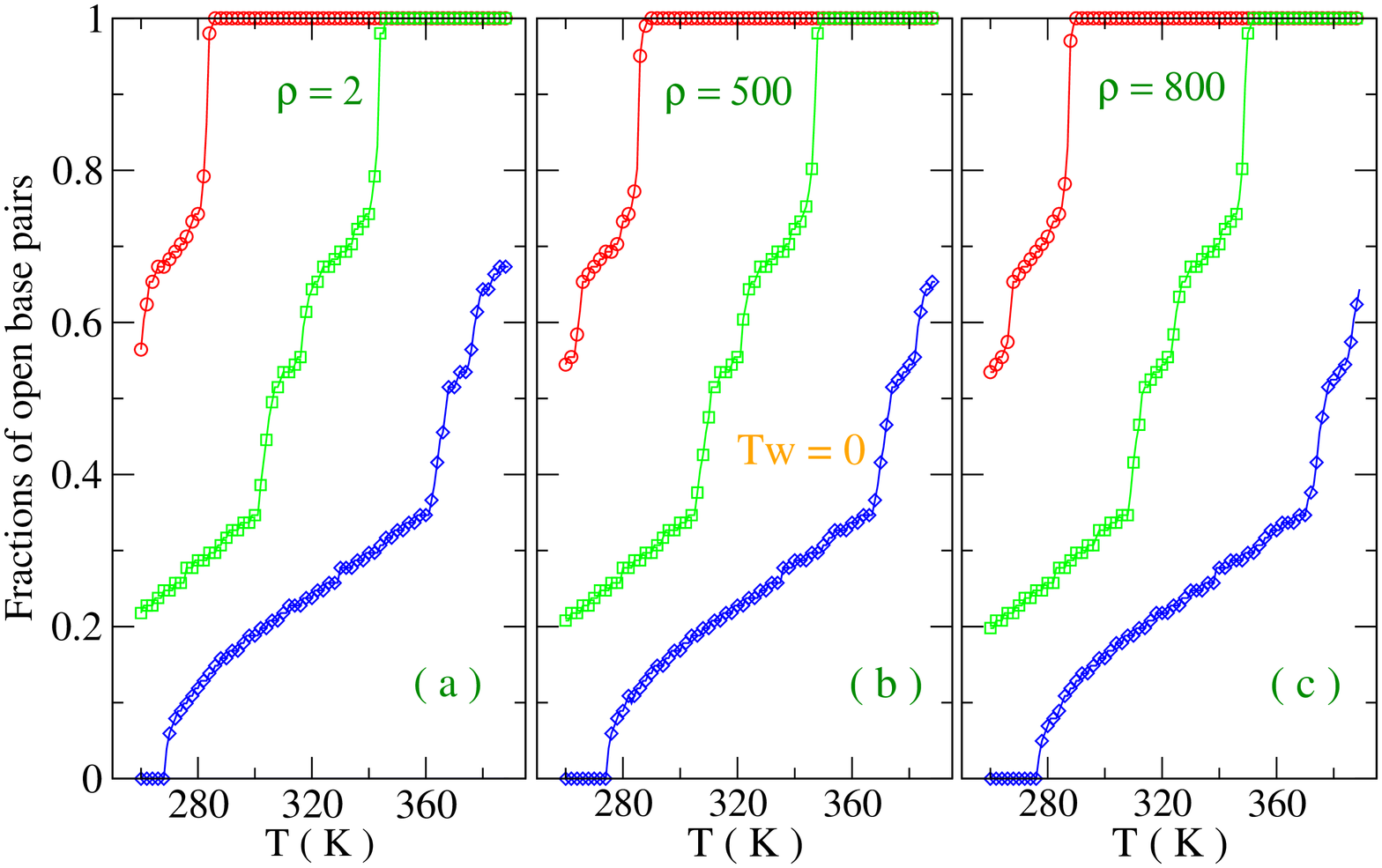}
\includegraphics[height=5.5cm,width=8.5cm,angle=-90]{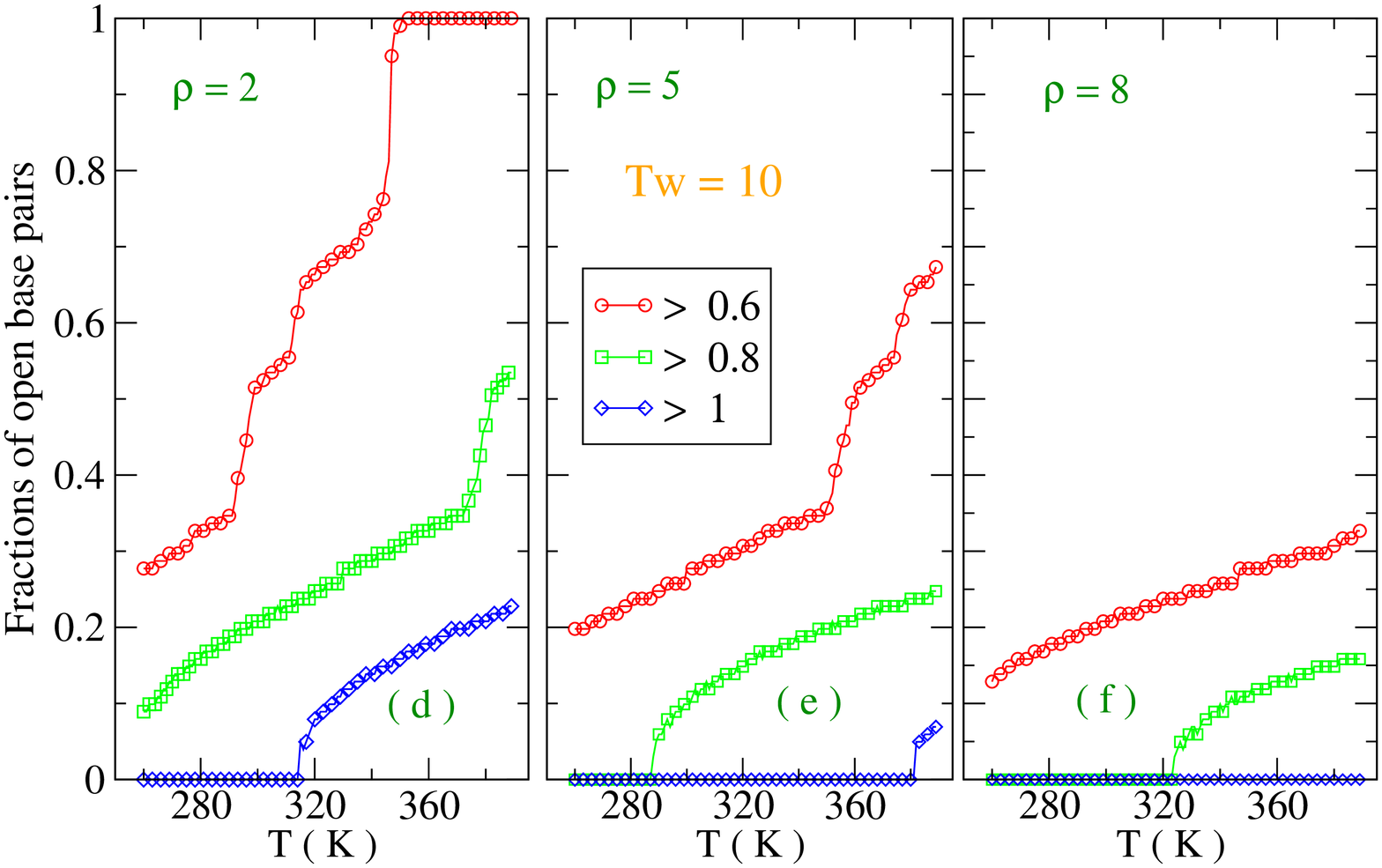}
\caption{\label{fig:2}(Color online) Fractions of average displacements larger than $\zeta=\,0.6$  (circles),  0.8 (squares),  1 ${\AA}$ (diamonds)  versus temperature in the  DPB ladder model ((a) - (c)) and in the equilibrium twist conformation ((d) - (f)). Three anharmonic stacking $\rho$ are assumed both in the left and in the right panel. In {(e)} and {(f)}, $\rho$ is smaller by a factor $100$ than in (b) and (c) respectively. }
\end{figure}

\section{Anharmonic Stacking versus Twist}

The computational method is here applied to investigate the interplay between twisting and stacking anharmonicity also in view of the special role given to the latter in the DPB model where a finite $\rho$  induces a sharp denaturation transition driven by sizeable melting entropy \cite{pey2}.
In Fig.~\ref{fig:2} the melting profiles are plotted \emph{both} for the DPB model with zero twist (left panel) and \emph{for} the equilibrium twist conformation (right panel). Several $\rho$ values are assumed in both cases.
In the $Tw=\,0$ system, even very large $\rho$ produce scant variations in the denaturation patterns pointing to a substantial irrelevance of the $\rho$ driven anharmonicity for the DPB ladder model.
Quite different is the physical picture emerging from the right panel regarding the $(Tw)_{eq}$ conformation: even slight enhancements over the $\rho=\,2$ value \emph{ i)} shift upwards along $T$ the opening of the average base pair displacements and \emph{ii)} flatten the melting profiles suggesting that the denaturation becomes more gradual.  Note that $\rho$ values in Figs.~\ref{fig:2}(e),(f) are two orders of magnitude smaller than in Figs.~\ref{fig:2}(b),(c) respectively. The $\rho=\,8$ plot says that, even at $T \sim \,400K$, there are no average displacements larger than $1 {\AA}$ while only $35\%$ are larger than $0.6 {\AA}$. Thus it is the twisting that switches on the $\rho$ effect. The latter induces those cooperative interactions  along the molecule backbone which are peculiar of the fluctuational openings. Accordingly, the anharmonic stacking renders the double helix flexible hence it does increase the molecule resilience against the whole thermal disruption of the hydrogen bonds.
In this sense, anharmonicity is a stabilizing factor for the double helix. Modeling DNA by path integrals offers the advantage to include in the computation a great number of molecule configurations meanwhile accounting for those fluctuational effects which are key to the molecule dynamics, mainly in short fragments.


\begin{references}


\bibitem{bates}
Bates A D, Maxwell A (2009) \emph{DNA Topology} (Oxford University Press, Oxford, UK)


\bibitem{pey2}
Dauxois T, Peyrard M and Bishop A R  (1993) \emph{Phys. Rev. E} \textbf{47} R44-R47

\bibitem{io09}
Zoli M (2009)  \emph{Phys. Rev. E} \textbf{79} 041927; Zoli M (2010) \emph{Phys. Rev. E } \textbf{81} 051910


\bibitem{io11}
Zoli M (2011)  \emph{J. Chem. Phys.} \textbf{135} 115101; Zoli M (2011) \emph{Eur. Phys. J. E} {\bf 34} 68


\bibitem{fehi}
Feynman R P and Hibbs A R (1965) {\it Quantum Mechanics and Path Integrals}, (Mc Graw-Hill, New York)


\end{references}
\end{document}